\documentclass[aps,prb,twocolumn,showpacs,groupedaddress]{revtex4-1}

\def\comment#1{\bigskip\hrule\smallskip#1\smallskip\hrule\bigskip}   
\usepackage{amsmath}
\usepackage{graphicx}
\usepackage{dcolumn}
\usepackage{ulem}
\usepackage{bm}
\usepackage{color}

\begin{document}

\title{Quantum Monte Carlo study of spin-polarized deuterium}

\author{I. Be\v{s}li{\'c}$^{1,2}$, L. Vranje\v{s} Marki{\'c}$^{2}$, J. Casulleras$^{1}$, 
J. Boronat$^{1}$}

\affiliation{$^1$ Departament de F\'\i sica i Enginyeria Nuclear, Campus Nord
B4-B5, Universitat Polit\`ecnica de Catalunya, E-08034 Barcelona, Spain \\
$^2$ Faculty of Science, University of Split, HR-21000 Split, Croatia}
\def\comment#1{\bigskip\hrule\smallskip#1\smallskip\hrule\bigskip}   
\def\qr{{\bf r}}

\date{\today}

\begin{abstract}
The ground-state properties of spin-polarized deuterium (D$\downarrow$) at zero temperature are obtained by means of diffusion Monte Carlo calculations within the fixed-node approximation. Three D$\downarrow$ species have been investigated (D$\downarrow_1$, D$\downarrow_2$, D$\downarrow_3$), corresponding respectively to one, two, and three equally occupied nuclear-spin states. The influence of the backflow correlations on the ground-state energy of the systems is explored. The equations of state of liquid D$\downarrow_2$ and D$\downarrow_3$ are obtained and compared with the ones obtained in previous approximate predictions. The density and pressure at which D$\downarrow_1$ experiences a gas-liquid transition at \textit{T}=0 are obtained.
\end{abstract}


\maketitle

\section{Introduction}

As the simplest element in nature, hydrogen has been investigated theoretically and
experimentally from the very first beginnings of quantum theory. Since hydrogen appears in
three isotopic forms (hydrogen, deuterium, and tritium), it offers even more interesting
scientific investigation possibilities. The interest of the scientific community in the
study of electron spin-polarized hydrogen (H$\downarrow$) and its isotopes, spin-polarized
deuterium (D$\downarrow$) and spin-polarized tritium (T$\downarrow$), began after Kolos
and Wolniewicz (KW)  calculated very precisely the triplet pair potential $b$$^3\Sigma_u^+$
in 1965.\cite{kolos} The enthusiasm for the study of electron spin-polarized hydrogen and
its isotopes originated from the expectations to explore even more extreme quantum matter
than helium isotopes.\cite{mil-nos-par,stw-nos,mil-nos} Such expectations were grounded on
the extremely weak attraction of the triplet pair potential  $b$$^3\Sigma_u^+$ through
which two H$\downarrow$ (D$\downarrow$ or T$\downarrow$) atoms interact, and on their even
lighter masses compared to  helium isotopes. In addition, it was shown by Freed
\cite{freed} in 1980 that, within the Born-Oppenheimer approximation in the spin-aligned
electronic state, hydrogen nuclei behave as effective bosons, as well as tritium nuclei.

Stwalley and Nosanow \cite{stw-nos} had proposed in 1976 H$\downarrow$ as the most
promising candidate for achieving a Bose-Einstein condensate (BEC). This theoretical
prediction was an important impulse for the experimentalists because production of cold
samples always represented a huge experimental challenge. The extensive H$\downarrow$ study
started in Amsterdam in 1980 when Silvera and Walraven managed to stabilize a very dilute
gas of spin-polarized hydrogen.\cite{sil-wal,wal-sil} A long
experimental journey preceded the final realization of a BEC state in H$\downarrow$. In
1998, Fried \textit{et al.} \cite{fried1} managed to form a BEC state of H$\downarrow$
using an experimental setup with a wall-free confinement and a low evaporation rate.
All spin-polarized hydrogen isotopes are usually a mixture of hyperfine
states, which is important for confinement and stabilization of the system. In Ref.
\onlinecite{fried2}, Greytak \textit{et al.} concluded that it is not
possible to confine in a static magnetic trap the lowest two states, $a$ and $b$ 
(high-field seeking states, Fig. 1 in Ref. \onlinecite{fried2}), due to the impossibility of having a
maximum in the magnitude of the magnetic field in a source-free region.
Thus, stable states have to be sought among the $c$ and $d$ states (low-field seeking states, Fig. 1 in Ref. \onlinecite{fried2}), which in pure magnetic traps have a local minimum in the field. In the experiments described in Ref. \onlinecite{fried2}, the doubly polarized $d$ state was used, usually designated as H\,$\uparrow${\footnotesize \sout{$\uparrow$}}, in which both electron and
nuclear spins are polarized in the direction of the magnetic field. The second, crossed
arrow, refers specifically to the nuclear spin.

Those very important experimental achievements were accompanied by new theoretical work.
Recently, the ground-state properties of H$\downarrow$ have been investigated using the
diffusion Monte Carlo (DMC) method.\cite{hydrogen} By means of accurate microscopic
calculations it has been confirmed that H$\downarrow$ does not form a liquid phase at zero
temperature, and in addition the gas-solid phase transition was also examined. The ground-state
properties calculated with DMC are obtained using the triplet pair potential
$b$$^3\Sigma_u^+$, recently recalculated and extended to larger interparticle distances by
Jamieson \textit{et al.} (JDW).\cite{jamieson} The DMC results are in good agreement with
the conclusions previously obtained with different variational methods  by other
authors\cite{stw-nos,etters,lan-nie,ent-anl} concerning the gas phase of bulk
H$\downarrow$. In all of the theoretical studies, hyperfine interaction has
not been considered and no magnetic field has been included, so both H$\downarrow$ and
H$\uparrow$ refer to the same system. Nuclear spins are not explicitly labeled and different
hyperfine states are degenerate in that approach. The same approach has been
taken in all theoretical studies of bulk properties of other hydrogen isotopes as well.

The ground-state properties of tritium T$\downarrow$ have also been microscopically
studied. Due to its larger mass and the fact that T$\downarrow$ atoms obey Bose statistics,
it was predicted using variational theory that T$\downarrow$ forms a liquid at $T$=0.
\cite{mil-nos,etters,joudeh1} Those predictions have been recently confirmed using the DMC
method,\cite{tritium} and the densities at which the liquid-solid phase transition occurs
at $T$=0 have been also determined. As in the case of spin-polarized hydrogen,\cite{hydrogen}
the DMC simulations used the JDW interatomic potential. On the other hand, Blume \textit{et
al.} \cite{blume} pointed to T$\downarrow$ as a possible new BEC gas in an optical dipole
trap because of its very broad Feshbach resonance, which can be used to tune the
interaction potential.

From the experimental point of view, the exploration of D$\downarrow$ started almost
simultaneously with H$\downarrow$ experiments.\cite{sil-wal2} In the same group in which a
very dilute gas of spin-polarized hydrogen was stabilized,\cite{sil-wal,wal-sil} D$\downarrow$ was also the subject of experimentation. Contrary to the stabilization of H$\downarrow$, it was not possible to achieve stable 
D$\downarrow$ due to its adsorption on the $^4$He surface and its posterior recombination to form D$_2$. The
maximum of the achieved D$\downarrow$ density in that experiment was at least two orders of
magnitude lower than the one achieved for H$\downarrow$. Even though
D$\downarrow$ was then not experimentally stabilized, a lot of theoretical work 
was dedicated to this remarkable quantum system. 

Having in mind that D$\downarrow$ is a Fermi system with nuclear spin 1 and given the fact that
different D$\downarrow$ species are possible depending on how D$\downarrow$ atoms are
distributed with respect to the available nuclear-spin states (-1,~0,~+1), the ground-state
properties of this exciting system were studied in the
past.\cite{kro,pan-che,flynn,panoff,skjetne,panoffN} According to the best of
our knowledge, microscopic DMC calculations for spin-polarized deuterium have not been
performed yet, thus leaving the determination of the equation of state for D$\downarrow$
only at the variational level.

Previously, it was shown that when all three nuclear-spin states are equally occupied,
D$\downarrow_3$, the ground state is a liquid.\cite{pan-che} The ground state of the system
in which two nuclear-spin states are equally occupied, D$\downarrow_2$, was studied by
Panoff and Clark using variational Monte Carlo (VMC).\cite{panoff} They used an improved
wave function which included optimal Jastrow, triplet, and backflow correlations. The
negative energy per particle obtained at equilibrium density $\rho_0=0.004 $ \AA$^{-3}$
revealed that D$\downarrow_2$ also forms a self-bound liquid in the ground state. 

From the theoretical side, special attention was dedicated to
D$\downarrow_1$ because, as emphasized in Ref. \onlinecite{panoffN}, it has the
best chance of experimental realization due to its predicted gas nature. 
Important theoretical results regarding the lifetime of magnetically trapped D$\downarrow$
gas were given in the works of Koelman \textit{et al.}~\cite{koel1,koel2}  The population
dynamics of the hyperfine levels of atomic deuterium, presented in  Fig. 1 of Ref.
\onlinecite{koel1}, was investigated as a function of the applied magnetic field.  It was
shown~\cite{koel1,koel2} that a mixture of the low-field-seeking hyperfine $\delta$,
$\epsilon$, and $\zeta$ states, confined in a static minimum-\textit{B}-field trap, will
decay rapidly, due to spin exchange, towards the doubly polarized gas
D\,$\uparrow${\footnotesize \sout{$\uparrow$}} of only $\zeta$-state atoms. It was also
shown that D\,$\uparrow${\footnotesize \sout{$\uparrow$}} stability should grow with
decreasing temperature. In that way D\,$\uparrow${\footnotesize \sout{$\uparrow$}} was
characterized as the most stable \textit{B}-field-trappable spin-polarized system.   
Since in the theoretical studies of D$\downarrow_1$ magnetic field is not considered, the
direction of the spins is not specified, so the obtained results refer to both
D\,$\uparrow${\footnotesize \sout{$\uparrow$}} and D\,$\downarrow${\footnotesize
\sout{$\downarrow$}}. However, after the results of Koelman \textit{et
al.}\cite{koel1,koel2} it became clear that D\,$\uparrow${\footnotesize \sout{$\uparrow$}}
is the version of D$\downarrow_1$ (Ref. \onlinecite{dave}) which is the most likely to be 
experimentally achieved. In 1995 Hayden and Hardy studied extensively atomic hydrogen and deuterium
mixtures confined by liquid-helium-coated-walls.\cite{hayden} The technique used in their experiments 
enabled obtaining the information about the two atomic densities simultaneously. In addition, recently, 
magnetic trapping of the low-field-seeking deuterium atoms after multistage Zeeman deceleration was achieved,\cite{Wiedekher} opening prospects for the experimental study of this system.

The nature of the ground state of D$\downarrow_1$ was not fully resolved in Ref.
\onlinecite{panoff} because of the obtained positive variational energy per particle, even
when the refinements to the ground-state trial wave function were included in the
description of the system. Therefore, they could not predict with certainty the
zero-temperature phase of D$\downarrow_1$ and concluded that D$\downarrow_1$ may remain in
the gaseous state down to absolute zero, leaving open the possibility that D$\downarrow_1$
could liquefy under a very slight pressure. Their results have been qualitatively confirmed
recently by Skjetne and \O stgaard with the Silvera interaction potential.\cite{skjetne} In
our recent VMC study of bulk D$\downarrow_1$ we investigated the influence of the
interaction potential on the D$\downarrow_1$ equation of state, and we discussed the
liquid-gas coexistence region.\cite{BVMCB} Our variational calculations showed that a
gas-liquid transition occurs at extremely low density of the gas ($\sim$10$^{-5}$ \AA$^{-3}$) and very low pressure ($\sim$0.0008 bar).

In the present work, we present results obtained using the diffusion Monte Carlo method for
the three spin-polarized deuterium species, D$\downarrow_1$, D$\downarrow_2$, and
D$\downarrow_3$, at zero temperature. The sign problem due to their Fermi statistics is
treated within the fixed-node (FN) approximation. This approximation is one of the most
accurate theoretical methods for the prediction of the ground-state properties of Fermi
systems, especially in cases in which the nodal surface of the trial wave function is very
close to the exact one. Interaction between D$\downarrow$ atoms is modeled using the
newest JDW triplet pair potential $b$$^3\Sigma_u^+$. This interaction is then smoothly
connected to the long-range behavior stated in Ref. \onlinecite{yan}. In addition to the
microscopic results of the energetic and structural properties of the D$\downarrow_1$,
D$\downarrow_2$, and D$\downarrow_3$ bulk systems at $T$=0, we comment also on the influence
of the backflow correlations on the ground-state energy at the DMC level. The obtained
equilibrium densities for D$\downarrow_3$ and D$\downarrow_2$ liquids are compared with
those determined in previous approximate descriptions. The gas-liquid phase transition of
D$\downarrow_1$ at $T$=0 is explored using the DMC method.

In Sec. II, the DMC method and the trial wave function used for importance sampling are
briefly described. The results obtained for the three spin-polarized deuterium species,
D$\downarrow_1$, D$\downarrow_2$, and D$\downarrow_3$, including the equations of state as
well as their structural properties, are presented in Sec. III. Finally, the
main conclusions of this work are summarized in Sec. IV.

\section{Method}
\label{sec:method}

The aim of the diffusion Monte Carlo method is to solve stochastically the Schr\"odinger equation written in imaginary time,
\begin{equation}
-\hbar \frac{\partial \Psi(\bm{R},t)}{\partial t} = (H- E_{\text r}) \Psi(\bm{R},t) \ ,
\label{schro}
\end{equation}
for an $N$-particle Hamiltonian

\begin{equation}
 H = -\frac{\hbar^2}{2m} \sum_{i=1}^{N} \bm{\nabla}_i^2 + \sum_{i<j}^{N} V(r_{ij})  \ .
\label{hamilto}
\end{equation}
The constant $E_{\text r}$ in Eq. (\ref{schro}) acts as a reference energy, and $\bm{R} \equiv (\bm{r}_1,\ldots,\bm{r}_N)$ is known as a \textit{walker}, which collectively denotes the particle positions within the Monte Carlo methodology. Introducing importance sampling, the Schr\"odinger equation is rewritten in terms of the mixed distribution $\Phi(\bm{R},t)=\Psi(\bm{R},t)\psi(\bm{R})$, where $\psi(\bm{R})$ is a trial wave function. In the diffusion process, the mixed distribution $\Phi(\bm{R},t)$ is represented by a set of \textit{walkers}. The lowest-energy eigenfunction, not orthogonal to $\psi(\bm{R})$, survives in the limit $t \rightarrow \infty$, and then the sampling of the ground state for an $N$-body bosonic system is effectively achieved. Because of the sign problem which exists in the case of Fermi systems, $\Psi(\bm{R},t)\psi(\bm{R})$ is not always positive. To satisfy the condition $\Psi(\bm{R},t)\psi(\bm{R}) \geq  0$ one relies on the fixed-node approximation in which  $\Psi(\bm{R},t)$ and $\psi(\bm{R})$ have to change sign together, i.e., share the same nodes. Having this in mind, the fixed-node energy obtained in the $t \rightarrow \infty$ limit is an upper bound to the exact ground-state energy. 

The trial wave function used in the present simulations is the Jastrow-Slater model $\psi({\bf{R}})=\psi_A\psi_J=\psi_A\prod_{i<j}^{N}f(r_{ij})$, where $\psi_J$ is the Jastrow part of the trial wave function which describes the dynamical correlations induced by $V(r_{ij})$ and $\psi_A$ is the antisymmetric wave function which introduces the statistics of Fermi particles. The Schiff-Verlet (SV) \cite{sv} form 
\begin{equation}
\label{eq:sv}
f(r_{ij}) = \exp \left[- \frac{1}{2} \left( \frac{b}{ r_{ij}} \right)^{5} \right],
\end{equation}
with variational parameter $b$, is used to model the two-body correlations. The antisymmetric part of the trial wave function $\psi_A$ is modeled with a Slater determinant in the case of D$\downarrow_1$ and the product of two and three Slater determinants in the case of D$\downarrow_2$ and D$\downarrow_3$, respectively. Single-particle plane-wave orbitals are used in the Slater determinant, $\varphi_{\alpha_i}(\textbf{r}_j)=\exp(i\:\textbf{k}_{\alpha_i}\textbf{r}_j )$, which correspond to the exact wave function of the Fermi sea. 

As usual for bulk system simulations with a finite number of
particles, a size-dependence analysis has to be performed. We added to our
DMC results the standard tail corrections, i.e., corrections coming from
the finite size ($L$) of the simulation box, 
\begin{eqnarray}
(E/N)_{\rm t} (\rho) & =  & 2 \pi \rho \int_{L/2}^{\infty} dr \ r^2 V(r) \\
\nonumber  
 & & - \frac{\hbar^2}{m}
\, 2 \pi \rho \int_{L/2}^{\infty} dr \ 
r^2 \left( \frac{2}{r} \, u^{\prime}(r) + u^{\prime \prime}(r) \right ) \ ,
\label{pot.tail}
\end{eqnarray}
assuming $g(r)=1$ for $r>L/2$, with $u(r)=-0.5 (b/r)^5$.
We also add the Fermi correction for the ground-state energy of the system. The Fermi correction is obtained as
the difference between the ground-state energy per particle of a free Fermi
gas [$\frac{3}{5} \frac{\hbar^2}{2m}\left( \frac{6 \pi^2 \rho}{\nu}
\right)^{\frac{2}{3}}$], with $\nu$ being the spin degeneracy, and the result
obtained by summing the discrete contributions of wave vectors
($\frac{\hbar^2k^2}{2m}$) used in our finite $N$-particle simulation.

As shown in Ref. \onlinecite {BVMCB}, it is enough to use 33 particles to simulate bulk D$\downarrow_1$. The same values of the optimal variational parameter $b$ reported in Ref. \onlinecite{BVMCB} are used in the present diffusion Monte Carlo calculations of D$\downarrow_1$ for the investigated density range, from $0.00009$ to $0.00634$ \AA$^{-3}$.
In the case of D$\downarrow_2$ and a density range from $0.00282$ to $0.00634$ \AA$^{-3}$, the optimal value of parameter $b$ slightly increases from $3.89$ to $3.97$  \AA. The detailed size-dependence analysis for D$\downarrow_2$ at density $\rho=0.00493 $ \AA$^{-3}$ is given in Table \ref{tab:sizeD2}. As shown in Table \ref{tab:sizeD2}, the minimum value of the Fermi correction is produced for the 66-particle system, and thus we decided to use this number of particles in our study.

\begin{table}[h!]
\begin{center} 
\begin{tabular}{ccccc} 
\\ \hline
\hline
~~~~~~~~ &  ~~~~ $E/N$ ~~~~~ & ~~~~ $E/N$  ~~~~~ & ~~~~Fermi~~~~~     & ~~~ $E/N$  ~~~~\\
~~~~$N$~~~~ &  ~~~~ VMC ~~~~~ & ~~~~ tail ~~~~~ & ~~~~correction~~   & ~~~ total~~~~
\\ \hline
38        &   0.81(1)         &    -0.305       &      0.079         &   0.58(1) \\
54        &   0.84(1)         &    -0.223       &      0.075         &   0.69(1) \\
66        &   0.84(1)         &    -0.188       &      0.01          &   0.66(1) \\
114       &   0.83(1)         &    -0.119       &     -0.031         &   0.68(1) 
\\ \hline
\hline
\end{tabular} 
\end{center}
\caption{Energy per particle of D$\downarrow_2$ (in K) as a function of the number of atoms included in the simulation for the density $\rho=0.00493 $ \AA$^{-3}$. Results are obtained with the VMC method and the JDW interatomic potential.\cite{jamieson} The numbers in parentheses are the statistical errors.}
\label{tab:sizeD2}
\end{table}

In the case of D$\downarrow_3$ and $N=57$ the value $b=3.93 $ \AA~ minimizes the energy per particle at density $\rho=0.00352 $ \AA$^{-3}$. This value coincides with the one used by  Panoff and Clark in their VMC study.\cite{panoff} To check whether this value of parameter $b$ does really optimize the energy, we have performed additional optimizations with 99 particles. In the density range from $0.00282$ to $0.00634$ \AA$^{-3}$ the optimal value of parameter $b$ does not change significantly. Detailed size-dependence analysis in the case of density $\rho=0.00282 $ \AA$^{-3}$  is given in Table \ref{tab:sizeD3}. The minimum value of the Fermi correction is obtained for $N=99$ particles, which we decided to use in the D$\downarrow_3$ DMC calculations.

\begin{table}[h!]
\begin{center} 
\begin{tabular}{ccccc} 
\\ \hline
\hline
~~~~~~~~ &  ~~~~ $E/N$ ~~~~~ & ~~~~ $E/N$  ~~~~~ & ~~~~Fermi~~~~~     & ~~~ $E/N$  ~~~~\\
~~~~$N$~~~~ &  ~~~~ VMC ~~~~~ & ~~~~ tail ~~~~~ & ~~~~correction~~   & ~~~ total~~~~
\\ \hline
57        &   0.071(9)         &    -0.077       &      0.041         &   0.035(9) \\
99        &   0.101(9)         &    -0.047       &      0.005         &   0.060(9) \\
171       &   0.081(5)         &    -0.028       &     -0.017         &   0.036(5)  
\\ \hline
\hline
\end{tabular} 
\end{center}
\caption{Energy per particle for D$\downarrow_3$ as a function of the number of atoms included in the simulation (in K) at the density $\rho=0.00282 $ \AA$^{-3}$. Results are obtained with the VMC method and the JDW interatomic potential.\cite{jamieson} The numbers in parentheses are the statistical errors.}
\label{tab:sizeD3}
\end{table} 

Since the results of the FN approximation depend on the \textit{quality} of
the trial wave function $\psi$, we improved the trial wave function by
introducing momentum-dependent correlations in the antisymmetric trial wave
function. These backflow correlations have been modeled in a similar way
as in the work by Panoff and Clark,\cite{panoff} but omitting the
long-range term ($\lambda_B^\prime/ r^3$), i.e., in the following way:
\begin{equation}
\label{eq:back}
\textbf{\~{r$_j$}} = \textbf{r$_j$}+\lambda_B \sum_{k\neq j}  \exp\left[ -\left( \dfrac{r_{jk}-r_B}{\omega_B} \right)^{2} \right]  (\textbf{r$_j$}-\textbf{r$_k$}),
\end{equation}
where $\lambda_B$, $r_B$, and $\omega_B$ are variational parameters.

At the variational Monte Carlo level the introduction of backflow correlations in the case of D$\downarrow_1$
could not improve the results because the optimization led to $\lambda_B  \rightarrow 0$ and the decrease of the energy was practically zero. The variational parameters that minimized the energy per particle of bulk D$\downarrow_2$ are $\lambda_B$=0.14, $r_B$=2.33 \AA, and $\omega_B$=1.68 \AA, while those of bulk D$\downarrow_3$ are $\lambda_B$=0.56, $r_B$=1.28 \AA, and $\omega_B$=2.53 \AA. For the three species we have minimized the variational parameters of the backflow correlations at density 0.00352 \AA$^{-3}$, a value which is close to the equilibrium density obtained with VMC when backflow correlations are not included in the wave function.

For all three spin-polarized deuterium species we used a DMC method accurate to second order in the time step $\Delta t$.\cite{time} In order to reduce any systematic bias coming from the time step and the mean number of walkers used in simulations, we investigated carefully possible dependences. Typical values of the number of walkers and the time step are 400 and 1$\times$10$^{-4}$ to 3$\times$10$^{-4}$ K$^{-1}$, respectively.

\begin{figure}[hbt]
\includegraphics[width=0.7\columnwidth]{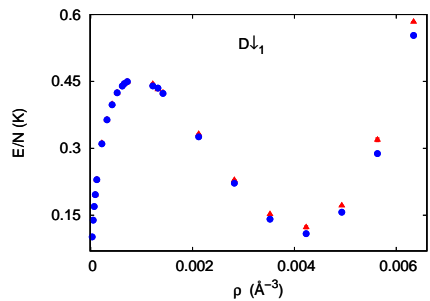}
\caption[]{\label{FIG:D1_final}
Energy per particle of D$\downarrow_1$ without backflow correlations (solid triangles) and with backflow correlations (solid circles) as a function of the density $\rho$. The error bars of the DMC energies are smaller than the size of the symbols.}
\end{figure}

\section{Results}
\label{sec:results}

\subsection{ D$\downarrow_1$ }

Bulk D$\downarrow_1$ is studied in the density range from $0.00003$ to $0.00634$ \AA$^{-3}$; the DMC energies are plotted in Fig. \ref{FIG:D1_final}. Since at the VMC level the introduction of backflow correlations in the case of D$\downarrow_1$ could not improve the results, we tried additional minimizations of the backflow parameters at the DMC level. For $\rho=0.00423$ \AA$^{-3}$, which is a density close to the value at which the energy shows a minimum, we performed additional DMC calculations in which we changed the backflow parameters $\lambda_B$, $r_B$, and $\omega_B$. In that way, we tried to explore the \textit{quality} of the nodes of our trial wave function. This search showed that the energy per particle slightly decreases if the DMC calculations are performed with values $\lambda_B$=0.14, $r_B$=2.43 \AA, and $\omega_B$=1.52 \AA. The decrease of the energy is practically negligible in the region of low density, while for larger
densities the decrease in energy increases to $\sim$13\%.

We fitted our DMC results in the density range from $0.00282$ to $0.00634$ \AA$^{-3}$ using the polynomial form ($e \equiv E/N$)

\begin{equation}
e(\rho)  = e_0+B\left(\frac{\rho-\rho_0}{\rho_0}\right)^2+C\left(\frac{\rho-\rho_0}{\rho_0}\right)^3  ,
\label{equat}
\end{equation}
with $\rho_0$ and $e_0$ being, respectively, the density and the energy per particle at the minimum. For several densities, we report the total and kinetic energy per particle in Table \ref{tab:D1}. The kinetic energy is calculated as the difference between the total energy and the pure estimation of the potential energy.\cite{pure} In this way, the bias coming from the choice of the trial wave function used in the simulations is removed.

\begin{table}
\begin{center}
\begin{ruledtabular} 
\begin{tabular}{lrrrr}         $\rho$ (\AA $^{-3}$) & 
~$E/N$ (K)    & ~~$T/N$ (K)    & ~~$P$ (bars)   & $c$ (m/s)     \\ \hline 
0.00423             &    0.109(1)   &   5.75(1)   &     0.01(1)       &     106(4)       \\
0.00493             &    0.157(2)   &   6.94(1)   &     0.42(2)       &     157(6)       \\
0.00563             &    0.288(2)   &   8.27(1)   &     1.21(6)       &    210(11)       \\
0.00634             &    0.553(2)   &   9.75(1)   &     2.57(13)      &    267(17) 
\end{tabular} 
\end{ruledtabular}
\end{center}
\caption{Results for liquid D$\downarrow_1$ at different densities $\rho$, with backflow correlations included in the model: energy per particle $E/N$, kinetic energy per particle $T/N$, pressure $P$, and speed of sound $c$. Numbers in parentheses are the statistical errors.}
\label{tab:D1}
\end{table}

The pressure and the speed of sound can be determined from their thermodynamic definitions and the DMC equation of state. The pressure of the system is given by
\begin{equation}
P(\rho)=\rho^2 \left( \frac{\partial e}{\partial \rho} \right) \ ,
\label{pressure}
\end{equation}
and the speed of sound is given by 
\begin{equation} 
c^2(\rho)=\frac{1}{m} \left( \frac{\partial P}{\partial \rho} \right) \ .
\label{speed}
\end{equation} 
In Table \ref{tab:D1} we also report for several densities the pressure and the speed of sound.

\begin{figure}[hbt]
\includegraphics[width=0.7\columnwidth]{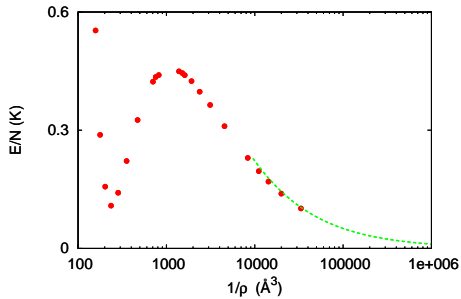}
\caption[]{\label{FIG:den_inv}
Energy per particle of D$\downarrow_1$ as a function of $1/\rho$ in logarithmic scale. The dashed line represents the universal equation of state of a Fermi gas in the region of very small densities.\cite{efimov}}
\end{figure}

When the backflow correlations were not included in the model, the fitting resulted in the best set of parameters $e_0=0.1246(4)$ K, $B=1.263(7)$ K, $C=0.83(1)$ K, and $\rho_0=0.004169(3) $ \AA$^{-3}$, and when the backflow correlations were included, the results were $e_0=0.1086(8)$ K, $B=1.31(2)$ K, $C=0.8(1)$ K, and $\rho_0=0.00420(3) $ \AA$^{-3}$. The backflow correlations move $\rho_0$ to slightly higher values and lower $e_0$ around 13\%. The comparison of our result [$e_0=0.1086(8)$ K and $\rho_0=0.00420(3) $ \AA$^{-3}$] with the best variational result of Panoff and Clark reported in Ref. \onlinecite{panoff} [$e_0=0.26(1)$ K and $\rho_0=0.004 $ \AA$^{-3}$] reveals also that DMC displaces $\rho_0$ to a higher density and lowers the energy per particle. 

As is expected, the decrease of the energy due to the diffusion process in the DMC method is more significant than the one caused by improving the trial wave function with backflow correlations. This is especially true in a fully polarized phase as D$\downarrow_1$ because no $s$-wave scattering is allowed.\cite{pandharipande}

\begin{figure}[hbt]
\includegraphics[width=0.7\columnwidth]{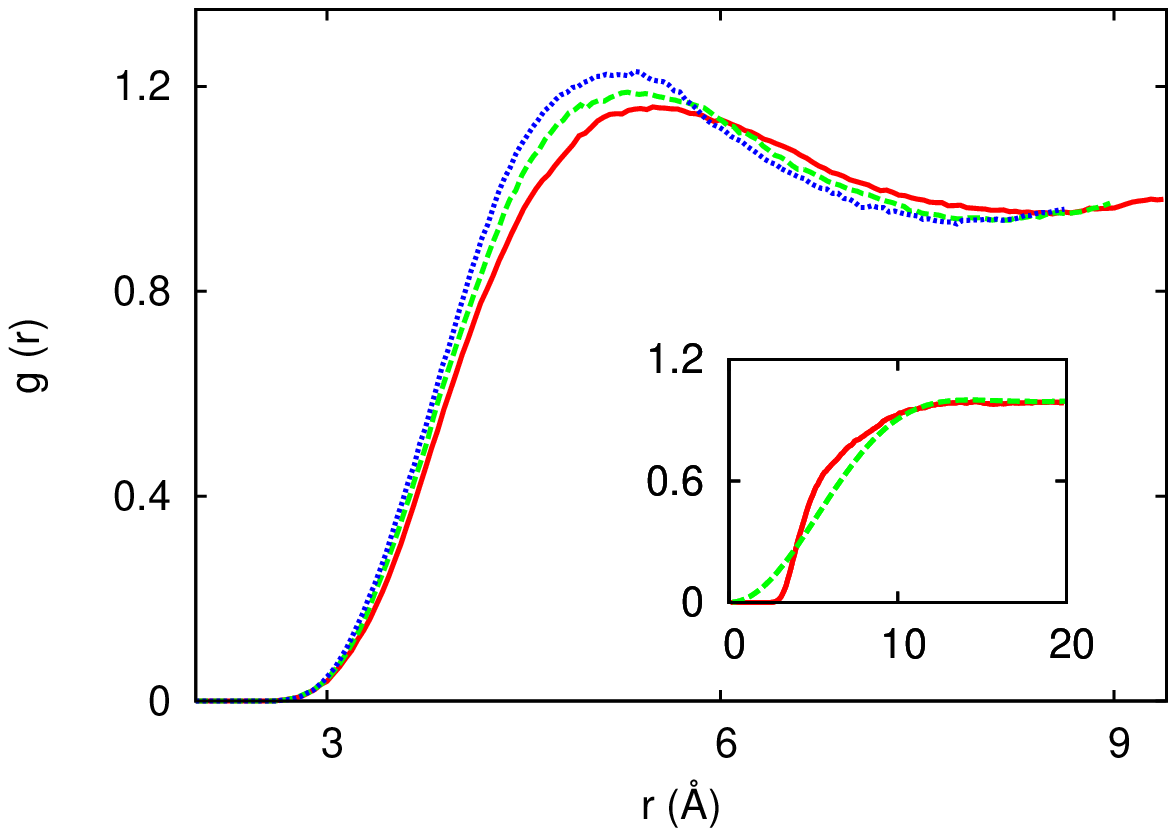}
\caption[]{\label{FIG:gr_D1}
Two-body radial distribution functions of D$\downarrow_1$. From bottom to top in the height of the main peak, the results correspond to densities 0.00493 \AA$^{-3}$ (solid line), 0.00563 \AA$^{-3}$ (dashed line), and 0.00634 \AA$^{-3}$ (dotted line). The inset shows $g(r)$ in the case of extremely low density $\rho=5.2\times10^{-4}$ \AA$^{-3}$ (solid line) and the free Fermi gas distribution function (dashed line).}
\end{figure}

The results for the energy per particle that we present within this work for D$\downarrow_1$ are in the liquid-gas coexistence region.\cite{mil-nos-par} In the literature, the liquid-gas coexistence region is defined as the region in which a first-order gas-liquid phase transition is possible at absolute zero. In order to construct a common tangent to the liquid and the gas equations of state, i.e., the double-tangent Maxwell construction, we plot in  Fig. \ref{FIG:den_inv} our results as a function of $1/\rho$, i.e., the volume per particle of the system. To proceed with the double-tangent Maxwell construction we included the universal equation of state of a Fermi gas in the region of very small densities.\cite{efimov} As one can see in Fig. \ref{FIG:den_inv}, DMC energies at densities $\rho < 10^{-4}$ \AA$^{-3}$ reproduce well this low-density expansion plotted as a dashed line. The presented results indicate that the gaseous state is the ground state of D$\downarrow_1$ and that D$\downarrow_1$ liquefies by applying just a very slight pressure at $T=0$. A first-order gas-liquid transition occurs at gas density $\rho= 1.48\times10 ^{-5}$ \AA$^{-3}$ [liquid density $\rho= 0.00421(1)$ \AA$^{-3}$] and at an extremely low pressure, $p\sim$ 9 $\times$10$^{-5}$ bar. With our former variational results, Ref. \onlinecite{BVMCB}, we predicted the transition at the gas density $\rho= 5.4\times10 ^{-5}$ \AA$^{-3}$ [liquid density $\rho= 0.00398(1)$ \AA$^{-3}$] and higher pressure, $p\sim$ 8 $\times$10$^{-4}$ bar. Both results of the transition imply that just the application of an extremely low pressure is enough to liquefy the gas.

Our conclusion regarding the liquid-gas coexistence region would be unaffected by the inclusion of a three-body
interaction potential.\cite{toennies} This interaction potential was used by Blume \textit{et al.} \cite{blume} in the study of the ground-state properties of small spin-polarized tritium clusters. In that work, Blume \textit{et al.} \cite{blume} showed that the inclusion of the three-body potential in the Hamiltonian results in a small increase of the ground state energy of the clusters.

\begin{figure}[hbt]
\includegraphics[width=0.7\columnwidth]{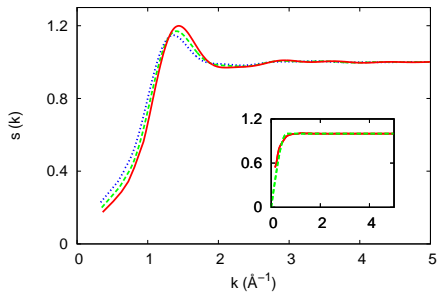}
\caption[]{\label{FIG:sk_D1}
Static structure function of D$\downarrow_1$. From bottom to top in the height of the main peak, the results correspond to densities 0.00493 \AA$^{-3}$ (dotted line), 0.00563 \AA$^{-3}$ (dashed line), and 0.00634 \AA$^{-3}$ (solid line). The inset shows $S(k)$ in the case of the extremely low density $\rho= 5.2 \times 10 ^{-4}$ \AA$^{-3}$ (solid line) and the free Fermi gas structure function (dashed line).}
\end{figure}

The ground-state structure properties of D$\downarrow_1$ are studied by calculating the two-body radial distribution function $g(r)$ and its Fourier transform, the static structure factor $S(k)$. In Figs. \ref{FIG:gr_D1} and \ref{FIG:sk_D1}, we show the DMC results obtained at different densities using the method of pure estimators.\cite{pure} As expected, the main peak of $g(r)$ shifts to shorter distances and its strength increases with the density. The more pronounced structure weakly emerges for the largest density included in the investigation, where a weak indication of the second peak formation can be recognized. The inset shows $g(r)$ for the density $\rho= 5.2 \times 10 ^{-4}$ \AA$^{-3}$ and the free Fermi gas distribution function (dashed line). Although the agreement between the two presented functions is not perfect, the DMC structural description of the very dilute regime of the system approaches the free Fermi gas except at very short distances, where the effect of the core of the interaction is evident.

A similar conclusion about the structure of the system can be derived from the
results of the static structure function $S(k)$ that we report in Fig.
\ref{FIG:sk_D1}. The reported results are the Fourier transforms of the
$g(r)$ functions except in the region of very small $k$, where we used
results obtained directly from the calculations,
\begin{equation}
S(k) = \frac{1}{N}\ \frac{\langle \Phi_0 | \rho_{\bf k}
\, \rho_{-{\bf
k}} | \Phi_0 \rangle}{\langle \Phi_0 | \Phi_0 \rangle} \ ,
\label{seq}
\end{equation}
with
\begin{equation}
 \rho_{\bf k} = \sum_{i=1}^{N} e^{i {\bf k} \cdot {\bf r}_i } \ .
\label{densi}
\end{equation}
In the very dilute regime,
shown in the inset, the obtained $S(k)$ reproduces very well the expected
$S(k)$ behavior of the free Fermi gas structure function.

\subsection{ D$\downarrow_2$ }

\begin{table}
\begin{center}
\begin{ruledtabular} 
\begin{tabular}{lrrrr}         $\rho$ (\AA $^{-3}$) & 
~$E/N$ (K)    & ~~$T/N$ (K)    & ~~$P$ (bars)   & $c$ (m/s)     \\ \hline 
0.00352             &    -0.040(3)   &   4.51(2)   &   -0.06(1)      &   70(3)        \\
0.00493             &     0.054(3)   &   6.84(2)   &    0.63(2)      &  167(7)        \\
0.00634             &     0.552(6)   &   9.68(4)   &    2.9(2)       &  276(15)
\end{tabular} 
\end{ruledtabular}
\end{center}
\caption{Results for liquid D$\downarrow_2$ at different densities $\rho$, with backflow correlations included in the model: energy per particle $E/N$, kinetic energy per particle $T/N$, pressure $P$, and speed of sound $c$. Numbers in parentheses are the statistical errors.}
\label{tab:D2}
\end{table}

\begin{figure}[hbt]
\includegraphics[width=0.7\columnwidth]{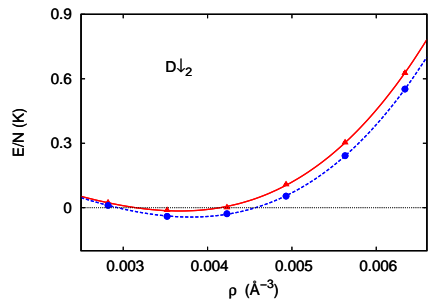}
\caption[]{\label{FIG:D2_final}
Energy per particle of liquid D$\downarrow_2$ without backflow correlations (solid triangles) and with backflow correlations (solid circles) as a function of the density $\rho$.  The solid and dashed lines correspond to fits to the DMC energies using Eq. (\protect\ref{equat}). The error bars of the DMC energies are smaller than the size of the symbols.}
\end{figure}

Bulk D$\downarrow_2$ is studied in the density range from $0.00282$ to $0.00634$ \AA$^{-3}$. In Table \ref{tab:D2}, we report the total and kinetic energies per particle for several densities. For all studied densities the DMC energies are plotted in Fig. \ref{FIG:D2_final}. The same analytical form  (\ref{equat}) that we used to fit the liquid part of the D$\downarrow_1$ data is used here to interpolate the equation of state of liquid D$\downarrow_2$. When the backflow correlations were not included, the fitting resulted with the best set of parameters $e_0=-0.015(3)$ K, $B=0.84(4)$ K, $C=0.52(9)$ K, and $\rho_0=0.00367(2) $ \AA$^{-3}$, and the equation of state (\ref{equat}) is shown as a solid line on top of the DMC data in Fig. \ref{FIG:D2_final}. When the backflow correlations were included in the model, the fitting resulted with the best set of parameters $e_0=-0.043(2)$ K, $B=0.96(2)$ K, $C=0.56(6)$ K, and $\rho_0=0.00381(5) $ \AA$^{-3}$, shown as a dashed line on top of the DMC data in Fig. \ref{FIG:D2_final}. In both cases the statistical uncertainties for the obtained fitting parameters are given as numbers in parentheses.

With the diffusion Monte Carlo method it is possible to obtain negative energies per particle for bulk D$\downarrow_2$ even without backflow correlations. We could not obtain negative energies per particle using the VMC method and a simple model of the trial wave function in which only the SV type of correlation is included, similar to what the authors in Ref. \onlinecite{panoff} noted. In the same work, Panoff and Clark reported $\rho_0=0.004$ \AA$^{-3}$ and  $e_0=-0.08(1)$ K when the trial wave function was improved with optimal Jastrow, triplet, and backflow correlations. Their reported result for the equilibrium energy per particle is lower than our best result, $e_0=-0.043(2)$ K. However, they used $N=54$ particles and did not take into account the Fermi correction (Table \ref{tab:sizeD2}) that amounts to $\sim$0.07 K. Including this correction, their VMC energy becomes $\sim$-0.01 K, clearly higher than our DMC result.

Concerning our results, one can see that the inclusion of backflow correlations moves the equilibrium density to a slightly higher value [from  $\rho_0=0.00367(2) $ \AA$^{-3}$ to $\rho_0=0.00381(5) $ \AA$^{-3}$] and expectedly lowers the equilibrium energy per particle [from $e_0=-0.015(3)$ K to $e_0=-0.043(2)$ K] . The value obtained in our DMC calculations for the equilibrium density $\rho_0=0.00381(5) $ \AA$^{-3}$ expressed in units of $\sigma^{-3}$ is $\rho_0=0.191~ \sigma^{-3}$ ($\sigma_H=3.6892$~\AA) which is slightly lower than the one obtained in the VMC calculations of Panoff and Clark,\cite{panoff} $\rho_0=0.2~ \sigma^{-3}$. 

Since liquid D$\downarrow_2$ resembles unpolarized liquid $^3$He, it is
useful to compare the equilibrium densities of both systems. Our result
for the equilibrium density of liquid D$\downarrow_2$, as well as the
result of Panoff and Clark, reveals a lower equilibrium density compared
to the one obtained in liquid $^3$He, $\rho_0=0.274~ \sigma^{-3}$
($\sigma_{He}=2.556$ \AA). A smaller difference was obtained in the
comparison of the equilibrium densities of Bose liquid T$\downarrow$ and
liquid $^4$He. Namely, the equilibrium density of liquid T$\downarrow$ in
terms of $\sigma_H$ is $\rho_0=0.375~ \sigma^{-3}$, and the equilibrium
density of liquid  $^4$He in terms of $\sigma_{He}$ is $\rho_0=0.365~
\sigma^{-3}$.

\begin{figure}[hbt]
\includegraphics[width=0.7\columnwidth]{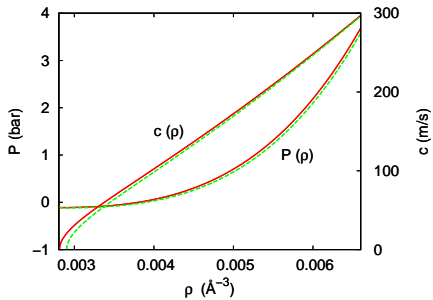}
\caption[]{\label{FIG:pre_spe_D2_D3}
Pressure and speed of sound of D$\downarrow_2$ (solid lines) and D$\downarrow_3$ (dashed lines) as a function of the density. Left (right) scale corresponds to pressure (speed of sound).}
\end{figure}

The extracted values for the pressure $P$ and the speed of sound $c$ for
several investigated densities are included in Table \ref{tab:D2}. In
addition, using Eqs. (\ref{pressure}) and (\ref{speed}) we obtained the
functions $P(\rho)$ and $c(\rho)$, which we show in Fig.
\ref{FIG:pre_spe_D2_D3}. The speed of sound becomes zero in liquid
D$\downarrow_2$ at the density $\rho_s=0.002813$ \AA$^{-3}=0.141
\sigma^{-3}$ and at a very small negative pressure, $P_{\text s}= -0.11(1)$
bar. In terms of $\sigma_H$, the spinodal density in liquid D$\downarrow_2$
is lower than in liquid $^3$He ($\rho_s=0.202 \sigma^{-3}$) and is even closer
to the equilibrium density than in liquid $^3$He. 

\begin{figure}[hbt]
\includegraphics[width=0.7\columnwidth]{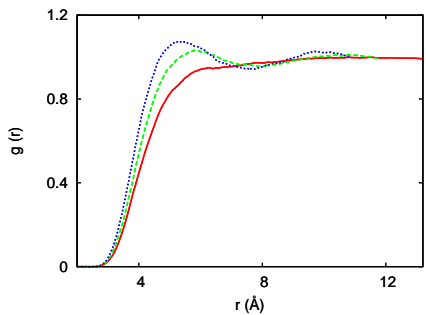}
\caption[]{\label{FIG:gr_same_D2}
Two-body radial distribution functions of liquid D$\downarrow_2$ for atoms with the same spin orientation. The results correspond to densities 0.00352 \AA$^{-3}$ (solid line), 0.00493 \AA$^{-3}$ (dashed line), and 0.00634 \AA$^{-3}$ (dotted line).}
\end{figure}

\begin{figure}[hbt]
\includegraphics[width=0.7\columnwidth]{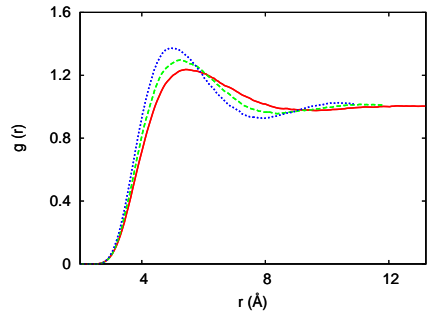}
\caption[]{\label{FIG:gr_dif_D2}
Two-body radial distribution functions of liquid D$\downarrow_2$ for atoms having different spin orientations. From bottom to top in the height of the main peak, the results correspond to densities 0.00352 \AA$^{-3}$ (solid line), 0.00493 \AA$^{-3}$ (dashed line), and 0.00634 \AA$^{-3}$ (dotted line).}
\end{figure}

\begin{figure}[hbt]
\includegraphics[width=0.7\columnwidth]{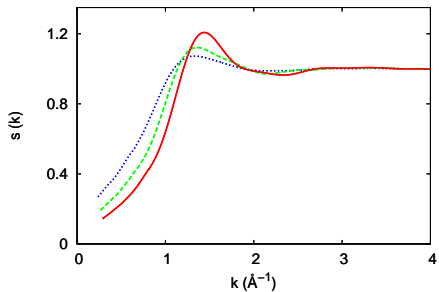}
\caption[]{\label{FIG:sk_D2}
Static structure function of D$\downarrow_2$. From bottom to top in the height of the main peak, the results correspond to densities 0.00352 \AA$^{-3}$ (dotted line), 0.00493 \AA$^{-3}$ (dashed line), and 0.00634 \AA$^{-3}$ (solid line).}
\end{figure}

The two-body radial distribution function is also obtained with pure estimators for liquid D$\downarrow_2$. We plot the DMC results for $g(r)$ of atoms having the same spin orientation in Fig. \ref{FIG:gr_same_D2} and of atoms having different spin orientations in Fig. \ref{FIG:gr_dif_D2}. In both cases, when  $\rho$ increases, the structure starts to become more pronounced, as can be seen in larger peaks and smaller first-neighbor distances. On the other hand, the comparison between the two-body radial distribution functions in Figs. \ref{FIG:gr_same_D2} and \ref{FIG:gr_dif_D2} reflects the spin-dependent difference which is a consequence of the Fermi statistics. Since, between atoms having the same spin orientation, repulsion is more effective, due to the Pauli principle, the main peak in Fig. \ref{FIG:gr_same_D2} at the density $\rho=0.00352$ \AA$^{-3}$ is practically not observed, while for the same density the main peak can be clearly localized in Fig. \ref{FIG:gr_dif_D2}. Also, due to the effective attraction between atoms having different spin orientations, the main peak at the density $\rho=0.00634$ \AA$^{-3}$ in Fig. \ref{FIG:gr_dif_D2} is significantly higher [$g(r)>1.2$] than the main peak at the same density in Fig. \ref{FIG:gr_same_D2} [$g(r)<1.2$]. A second peak emerging at larger $r$ can be recognized at $\rho=0.00634$ \AA$^{-3}$ for the cases with the same and different spin orientations of atoms. For the same densities, we report in Fig. \ref{FIG:sk_D2} the corresponding results for the total $S(k)$. Similar to D$\downarrow_1$ bulk, the main peak increases as the density increases.

\begin{figure}[hbt]
\includegraphics[width=0.7\columnwidth]{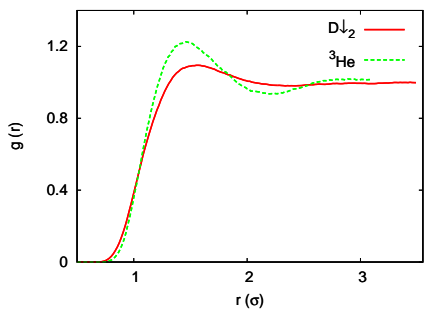}
\caption[]{\label{FIG:He3_D2_comp}
Two-body radial distribution functions of $^3$He ($r$ in $\sigma_{He}$) and D$\downarrow_2$ ($r$ in $\sigma_H$) liquids at the equilibrium densities.}
\end{figure}

In order to compare $^3$He and D$\downarrow_2$ liquids from the structural
perspective we have calculated the two-body radial distribution functions
at the equilibrium densities of $^3$He and D$\downarrow_2$. 
The FN-DMC calculation of liquid $^3$He has been carried out using the
HFD-B(HE) Aziz potential~\cite{aziz} and $N=66$ atoms. 
We plot both distribution functions in Fig. \ref{FIG:He3_D2_comp}, where $r$ is
expressed in terms of $\sigma_{He}$ and $\sigma_H$. It is clear from the
present results that the two liquids show different structure. The main
peak is significantly higher in the case of liquid $^3$He. Also, formation
of the second peak can be recognized in the case of liquid $^3$He, while
the second peak practically does not emerge in the case of liquid
D$\downarrow_2$. The obtained behavior is evidence of the stronger
interaction between $^3$He atoms.

\subsection{ D$\downarrow_3$ }

\begin{figure}[hbt]
\includegraphics[width=0.7\columnwidth]{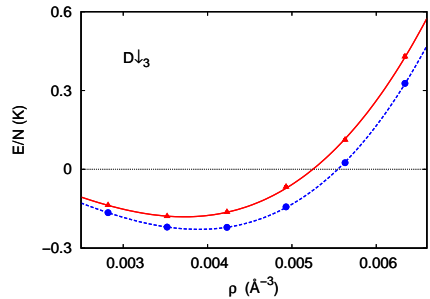}
\caption[]{\label{FIG:D3_final}
Energy per particle of liquid D$\downarrow_3$ without (solid triangles) and with (solid circles) backflow correlations as a function of the density $\rho$.  The solid and dashed lines correspond to fits to the DMC energies using Eq. (\protect\ref{equat}). The error bars of the DMC energies are smaller than the size of the symbols.}
\end{figure}

Liquid D$\downarrow_3$ is studied in the density range from $0.00282$ to $0.00634$ \AA$^{-3}$, and in Table \ref{tab:D3} we report for several densities the total and kinetic energies per particle. The DMC energies are plotted in Fig. \ref{FIG:D3_final}, and the equation of state is modeled by the analytical form  (\ref{equat}). When backflow correlations are not included in the trial wave function, the fitting gives a best set of parameters: $e_0=-0.181(2)$ K, $B=0.87(4)$ K, $C=0.52(7)$ K, and $\rho_0=0.00372(2) $ \AA$^{-3}$. The equation of state (\ref{equat}) is plotted as a solid line on top of the DMC data. When backflow correlations are incorporated, the best obtained set of parameters is $e_0=-0.229(1)$ K, $B=1.01(2)$ K, $C=0.64(5)$ K, and $\rho_0=0.00389(2) $ \AA$^{-3}$. The corresponding equation of state (\ref{equat}) is shown as a dashed line on top of the DMC data. In both cases, the statistical uncertainties for the obtained fitting parameters are given as numbers in parentheses.

\begin{table}
\begin{center}
\begin{ruledtabular} 
\begin{tabular}{lrrrr}         $\rho$ (\AA $^{-3}$) & 
~$E/N$ (K)    & ~~$T/N$ (K)    & ~~$P$ (bars)   & $c$ (m/s)     \\ \hline 
0.00352             &    -0.220(2)   &   4.34(2)      &   -0.08(1)      &   65(3)        \\
0.00493             &    -0.143(4)   &   6.60(4)      &    0.58(2)      &  164(5)        \\
0.00634             &     0.327(6)   &   9.34(4)      &    2.9(1)       &  274(11)
\end{tabular} 
\end{ruledtabular}
\end{center}
\caption{Results for liquid D$\downarrow_3$ at different densities $\rho$, with backflow correlations included in the model: energy per particle $E/N$, kinetic energy per particle $T/N$, pressure $P$, and speed of sound $c$. Numbers in parentheses are the statistical errors.}
\label{tab:D3}
\end{table}

Our DMC results show that the ground state of D$\downarrow_3$ is a liquid, even when backflow correlations are not incorporated in the trial wave function. Panoff and Clark\cite{panoff} reported negative VMC energies per particle for D$\downarrow_3$ in the case in which only the two-body correlations are used to model the trial wave function, indicating in that way that the ground state of the system is a liquid. We reproduced their VMC results using 57 atoms and the same variational parameter $b$ they used for the SV correlations, and we noticed that their conclusion about D$\downarrow_3$ ground state changes if the Fermi correction is added to the VMC results. In the density range from $0.00282$ to $0.00634$ \AA$^{-3}$, the Fermi correction for $N=57$ atoms increases from 0.04 to 0.07 K. Adding this correction to the VMC results, the energies become positive in the density range mentioned above, and one cannot predict a liquid phase for D$\downarrow_3$.

If we compare our DMC results in which the backflow correlations are absent with those in which the backflow correlations are included, it is clear that including the backflow correlations in the model shifts the equilibrium density to a slightly higher value [from  $\rho_0=0.00372(2) $ \AA$^{-3}$ to $\rho_0=0.00389(2) $ \AA$^{-3}$] and lowers the equilibrium energy [from $e_0=-0.181(2)$ K to  $e_0=-0.229(1)$ K], as it was already noticed for D$\downarrow_2$. In addition, the equilibrium density of D$\downarrow_3$ is always slightly higher than the equilibrium density of D$\downarrow_2$. The result reported for the D$\downarrow_3$ equilibrium density in Ref. \onlinecite{panoff} is slightly higher ($\rho_0=0.004$ \AA$^{-3}$) than the one we obtained with the DMC method. 

Even though the energy per particle at the equilibrium density $e_0=-0.229(1)$ K of D$\downarrow_3$ is lower than the one of D$\downarrow_2$
[$e_0=-0.043(2)$ K], it is still a small value, defining D$\downarrow_3$ as a weakly self-bound liquid, and also unique in the sense that it does not possess its helium analog.

\begin{figure}[hbt]
\includegraphics[width=0.7\columnwidth]{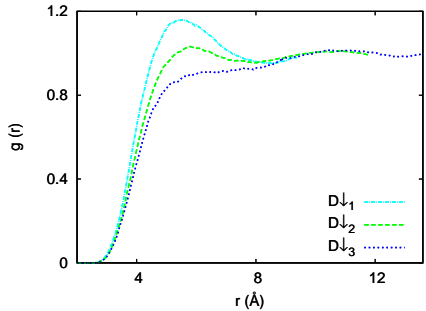}
\caption[]{\label{FIG:same_D}
Two-body radial distribution function for D$\downarrow_1$, D$\downarrow_2$, and D$\downarrow_3$ at the density 0.00493 \AA$^{-3}$ for atoms having the same spin orientation.}
\end{figure}

As in the case of D$\downarrow_2$, we used Eqs. (\ref{pressure}) and (\ref{speed}) to calculate the functions $P(\rho)$ and $c(\rho)$ for liquid D$\downarrow_3$. The extracted values for the pressure $P$ and the speed of sound $c$ for several densities are included in Table \ref{tab:D3}, while functions $P(\rho)$ and $c(\rho)$ for liquid D$\downarrow_3$ are added to Fig. \ref{FIG:pre_spe_D2_D3}, next to the results obtained for liquid D$\downarrow_2$. In liquid D$\downarrow_3$, the speed of sound becomes zero at the density $\rho_s=0.002903$ \AA$^{-3}=0.146 \sigma^{-3}$ and at a small negative pressure, $P_{\text s}= -0.12(1)$ bar. It is evident already from Fig. \ref{FIG:pre_spe_D2_D3} that there is a small difference between the spinodal densities in D$\downarrow_2$ and D$\downarrow_3$ liquids, even though the spinodal pressures are rather similar.
In addition, very similar values of the pressure and the speed of sound in D$\downarrow_2$ and D$\downarrow_3$ liquids are revealed from  Fig. \ref{FIG:pre_spe_D2_D3}, as well as from Tables \ref{tab:D2} and \ref{tab:D3} in the remaining region of investigated densities.

\begin{figure}[hbt]
\includegraphics[width=0.7\columnwidth]{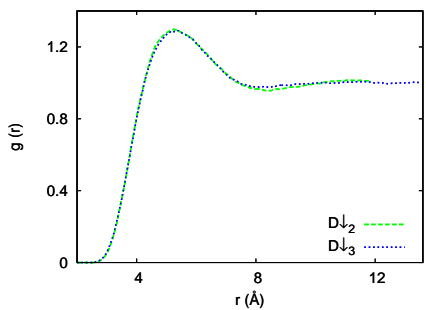}
\caption[]{\label{FIG:dif_D}
Two-body radial distribution function for D$\downarrow_2$ and D$\downarrow_3$ at the density 0.00493 \AA$^{-3}$ for atoms having different spin orientation.}
\end{figure}

The two-body distribution functions of D$\downarrow_3$ have also been calculated, and it is interesting to compare $g(r)$ of the three D$\downarrow$ species for atoms having the same spin orientation, as well as in D$\downarrow_2$ and D$\downarrow_3$ for atoms having different spin orientation. At the density 0.00493 \AA$^{-3}$ we show the spin-dependent $g(r)$ in Figs. \ref{FIG:same_D} and \ref{FIG:dif_D}.
The increase of the degeneration obviously produces the effect of ''density reduction'' in the case of $g(r)$ of atoms having the same spin orientation (Fig. \ref{FIG:same_D}). A similar effect is not noticed in the case of $g(r)$ of atoms having different spin orientation (Fig. \ref{FIG:dif_D}).

\begin{figure}[hbt]
\includegraphics[width=0.7\columnwidth]{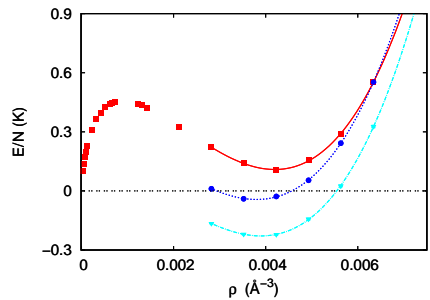}
\caption[]{\label{FIG:e_todos}
Energy per particle of D$\downarrow_1$ (solid  squares), D$\downarrow_2$ (solid  circles), and D$\downarrow_3$ (solid triangles) as a function of the density $\rho$, with backflow correlations included in the model. The lines correspond to fits to the DMC energies using Eq. (\protect\ref{equat}). The error bars of the DMC energies are smaller than the size of the symbols.}
\end{figure}

\section{Conclusions}

The ground-state properties of the three spin-polarized deuterium species
(D$\downarrow_1$, D$\downarrow_2$, D$\downarrow_3$) have been accurately
determined using the DMC method within the fixed-node approximation. The
accuracy of the DMC method and precise knowledge of the
D$\downarrow$-D$\downarrow$ interatomic potential allowed for a nearly
exact determination of the main properties of these systems. The best
obtained results for the energy per particle for all three D$\downarrow$
species are summarized in Fig. \ref{FIG:e_todos}. The energy ordering for
the three D$\downarrow$ species, close to the equilibrium densities, was
found to be $(E/N)_{D\downarrow_1}>(E/N)_{D\downarrow_2}>(E/N)_{D\downarrow_3}$ due to
the degeneracy, as was already pointed out in previous variational
descriptions of the systems. Interestingly, our results show that the
equations of state of D$\downarrow_1$ and D$\downarrow_2$ cross at pressure
$P=2.8(2)$ bars, pointing to a possible ferromagnetic transition from
D$\downarrow_2$ to D$\downarrow_1$.

Our study confirms previous variational predictions for the self-bound
quantum liquids D$\downarrow_2$ [$\rho_0=0.00381(5) $ \AA$^{-3}$] and
D$\downarrow_3$ [$\rho_0=0.00389(2) $ \AA$^{-3}$]. The spinodal densities
are determined for both liquids. We also discuss the ground state of
D$\downarrow_1$ and predict the gas density and the pressure at which
D$\downarrow_1$ liquefies at $T$=0, i.e., the conditions at which the
system undergoes a first-order gas-liquid phase transition.

Spin-polarized atomic deuterium is a paradigmatic example of the relevance
of quantum effects, mainly of the quantum statistics, on the nature of
condensed matter at very low temperatures. The interatomic potential does
not distinguish between the three D$\downarrow$ species because it
is dominated by the electronic structure. However, the occupation of the
nuclear spin states is able of producing different physical phases due to
the relative weight of the Fermi statistical correlations. When the Pauli
principle becomes more important, in the D$\downarrow_1$ case, the system
is no longer a liquid at zero pressure as happens in D$\downarrow_2$ and 
D$\downarrow_3$, but it is a gas. No such effect is observed in liquid
$^3$He in which the complete spin polarization does not change its
liquid character. It could be that atomic deuterium is the only physical system in
which this spin-degeneracy mechanism is able to modify zero-temperature phase
diagrams. We hope our work will stimulate further research of these extremely
quantum effects, from both theoretical and experimental sides.

\acknowledgments
We acknowledge partial financial support from DGI (Spain) Grant No.
FIS2011-25275, Generalitat de Catalunya Grant No. 2009SGR-1003, 
MSES (Croatia) Grant No. 177-1770508-0493, and Qatar National 
Research Fund Grant No. NPRP 5-674-1-114. These materials are based on
work financed by the Croatian Science Foundation, and I.B.
acknowledges the support. In addition, the resources of the Zagreb
University Computing Centre (Srce) and Croatian National Grid
Infrastructure (CRO NGI) were used, along with the resources of the HYBRID
cluster at the University of Split, Faculty of Science.

\end{document}